\documentclass[aps,prl,twocolumn,superscriptaddress,showpacs]{revtex4}
\usepackage{amsmath}
\usepackage{amstext}
\usepackage{amssymb}
\usepackage[dvips]{graphicx}

\newcommand{\beq}{\begin{equation}}
\newcommand{\eeq}{\end{equation}}
\newcommand{\ket}[1]{\left\vert#1\right\rangle}
\newcommand{\bra}[1]{\left\langle#1\right\vert}
\newcommand{\eps}{\varepsilon}

\begin{document}
       
\title{Robust and efficient generator of 
almost maximal multipartite entanglement}

\author{Davide Rossini}
\altaffiliation{Present address: International School
for Advanced Studies (SISSA), Via Beirut 2-4, I-34014 Trieste, Italy.}
\affiliation{NEST-CNR-INFM \& Scuola Normale Superiore,
  Piazza dei Cavalieri 7, I-56126 Pisa, Italy}

\author{Giuliano Benenti}
\affiliation{CNISM, CNR-INFM \& Center for Nonlinear and Complex systems,
Universit\`{a} degli Studi dell'Insubria, via Valleggio 11,
I-22100 Como, Italy}
\affiliation{Istituto Nazionale di Fisica Nucleare, Sezione 
di Milano, via Celoria 16, I-20133 Milano, Italy}

\date{\today}

\begin{abstract} 
Quantum chaotic maps can efficiently generate pseudo-random states 
carrying almost maximal multipartite entanglement, as characterized 
by the probability distribution of bipartite entanglement between 
all possible bipartitions of the system. We show that such multipartite 
entanglement is robust, in the sense that, when realistic 
noise is considered, distillable entanglement of bipartitions
remains almost maximal up to a noise strength that drops only polynomially 
with the number of qubits.
\end{abstract}

%03.67.Mn Entanglement production, characterization, and manipulation
%03.67.Lx Quantum computation
%03.67.-a Quantum information
%05.45.Mt Quantum chaos; semiclassical methods

\maketitle

Entanglement is not only the most intriguing feature of quantum mechanics,
but also a key resource in quantum information science~\cite{nielsen,qcbook}.
In particular, for quantum algorithms operating on pure states, multipartite
(many-qubit) entanglement is a necessary condition to achieve an
exponential speedup over classical computation~\cite{jozsa}.
The entanglement content of random pure quantum states is almost 
maximal~\cite{page93,zyczkowski,winter};
such states find applications in various quantum protocols, like
superdense coding of quantum states~\cite{hayden,winter},
remote state preparation~\cite{bennett}, and the construction of 
efficient data-hiding schemes~\cite{hayden2}. 
Moreover, it has been argued that random evolutions 
may be used to characterize the main aspects of noise sources affecting 
a quantum processor~\cite{saraceno}.

The preparation of a random state or, equivalently, the implementation
of a random unitary operator requires a number of elementary 
one- and two-qubit gates exponential in the number $n_q$ of qubits,
thus becoming rapidly unfeasible when increasing $n_q$.
On the other hand, pseudo-random states approximating to the desired
accuracy the entanglement properties of true random states may be 
generated efficiently, that is, polynomially in 
$n_q$~\cite{saraceno,weinstein,plenio}. In particular, quantum chaotic
maps are efficient generators of multipartite entanglement 
among the qubits, close to that expected for random 
states~\cite{caves,weinstein}.
A related crucial question is whether the generated entanglement 
is robust when taking into account unavoidable noise sources 
affecting a quantum computer, that in general turn pure states into 
mixtures, with a corresponding loss of quantum coherence
and entanglement content. 
In this paper we give a positive answer to this question. 

The number of measures needed to fully quantify multipartite entanglement
grows exponentially with the number of qubits.
Different measures capture various aspects of multipartite entanglement.  
Therefore, following Ref.~\cite{facchi}, we characterize multipartite
entanglement by means of a function rather than with a single measure:
we look at the probability density function of bipartite
entanglement between all possible bipartitions of the system.
For pure states the bipartite entanglement is the von Neumann entropy
of the reduced density matrix of one of the two subsystems:
%\beq
$E_{AB} (\ket{\psi} \bra{\psi})=
- {\rm Tr} \left[ \rho_A \log_2 \rho_A \right] \equiv S (\rho_{A}) \, ,$
%\eeq
where $\rho_A = {\rm Tr_B} \, (\ket{\psi} \bra{\psi})$, and $A, B$ denote
two subsystems made up of $n_A$ and $n_B$ qubits ($n_A + n_B = n_q$).
For sufficiently large systems ($N \equiv 2^{n_q} \gg 1$), 
it is reasonable to consider only balanced bipartitions, 
i.e., with $n_A = n_B$,
since the statistical weight of unbalanced ones % ($n_A\ll n_q$) 
becomes negligible~\cite{facchi}.
%; hereafter we will adopt this strategy.
If the probability density has a large mean value
$\langle E_{AB} \rangle \sim n_q$ ($\langle \, \cdot \, \rangle$ denotes 
the average over balanced bipartitions) and small relative standard 
deviation $\sigma_{AB}/\langle E_{AB}\rangle \ll 1$, we can conclude 
that genuine multipartite entanglement is almost maximal
(note that $E_{AB}$ is bounded within the interval $[0,n_q]$). 
This is the case for random states~\cite{facchi}.

\paragraph{The model.} The use of quantum chaos for efficient
and robust generation of pseudo-random states carrying large
multipartite entanglement is nicely illustrated by the example
of the quantum sawtooth map~\cite{benenti01}. 
This map is described by 
the unitary operator $\hat{U}$:
\beq
\ket{{\psi}_{t+1}} = \hat{U} \ket{\psi_t} =
e^{-iT\hat{n}^{2}/2} \, e^{ik(\hat{\theta} -\pi)^{2}/2} \ket{\psi_t} ,
\label{eq:quantmap}
\eeq
where $\hat{n} = -i \, \partial/\partial \theta$,
$[\hat{\theta},\hat{n}]=i$ (we set $\hbar=1$) and the discrete 
time $t$ measures the number of map iterations.
In the following we will always study map~\eqref{eq:quantmap}
on the torus $0 \leq \theta < 2 \pi$, $- \pi \leq p < \pi$,
where $p = T n$.
With an $n_q$-qubit quantum computer we are able to simulate the
quantum sawtooth map with $N = 2^{n_q}$ levels;
as a consequence, $\theta$ takes $N$ equidistant
values in the interval $0 \leq \theta < 2 \pi$, while
$n$ ranges from $-N/2$ to $N/2 -1$
(thus setting $T=2\pi/N$).
We are in the quantum chaos regime for map~\eqref{eq:quantmap}
when $K\equiv kT >0$ or $K<-4$; in particular, in this work 
we focus on the case $K=1.5$.

There exists an efficient quantum algorithm for simulating
the quantum sawtooth map~\cite{benenti01}. 
The crucial observation is that the operator $\hat{U}$ in
Eq.~\eqref{eq:quantmap} can be written as the product of two operators:
$\hat{U}_{k}= e^{ik(\hat{\theta}-\pi)^{2}/2}$
and $\hat{U}_{T}=e^{-iT\hat{n}^{2}/2}$,
that are diagonal in the $\theta$ and in the $n$ representation,
respectively. Therefore, the most convenient way to classically simulate
the map is based on the forward-backward fast Fourier 
transform between $\theta$ and $n$ representations, and requires
$O(N\log_2 N)$ operations per map iteration.
On the other hand, quantum computation exploits its capacity
of vastly parallelize the Fourier transform, thus requiring only 
$O((\log_2 N)^2)$ one- and two-qubit gates
to accomplish the same task~\cite{benenti01}. 
%More precisely, it needs $2n_q$ Hadamard gates and 
%$3n_q^2-n_q$ controlled-phase shift gates.
In brief, the resources required by the quantum 
computer to simulate the sawtooth map are 
only logarithmic in the system size $N$, thus admitting an
exponential speed up, as compared to any known classical computation.

\begin{figure}[!ht]
  \begin{center}
    \includegraphics[scale=0.33]{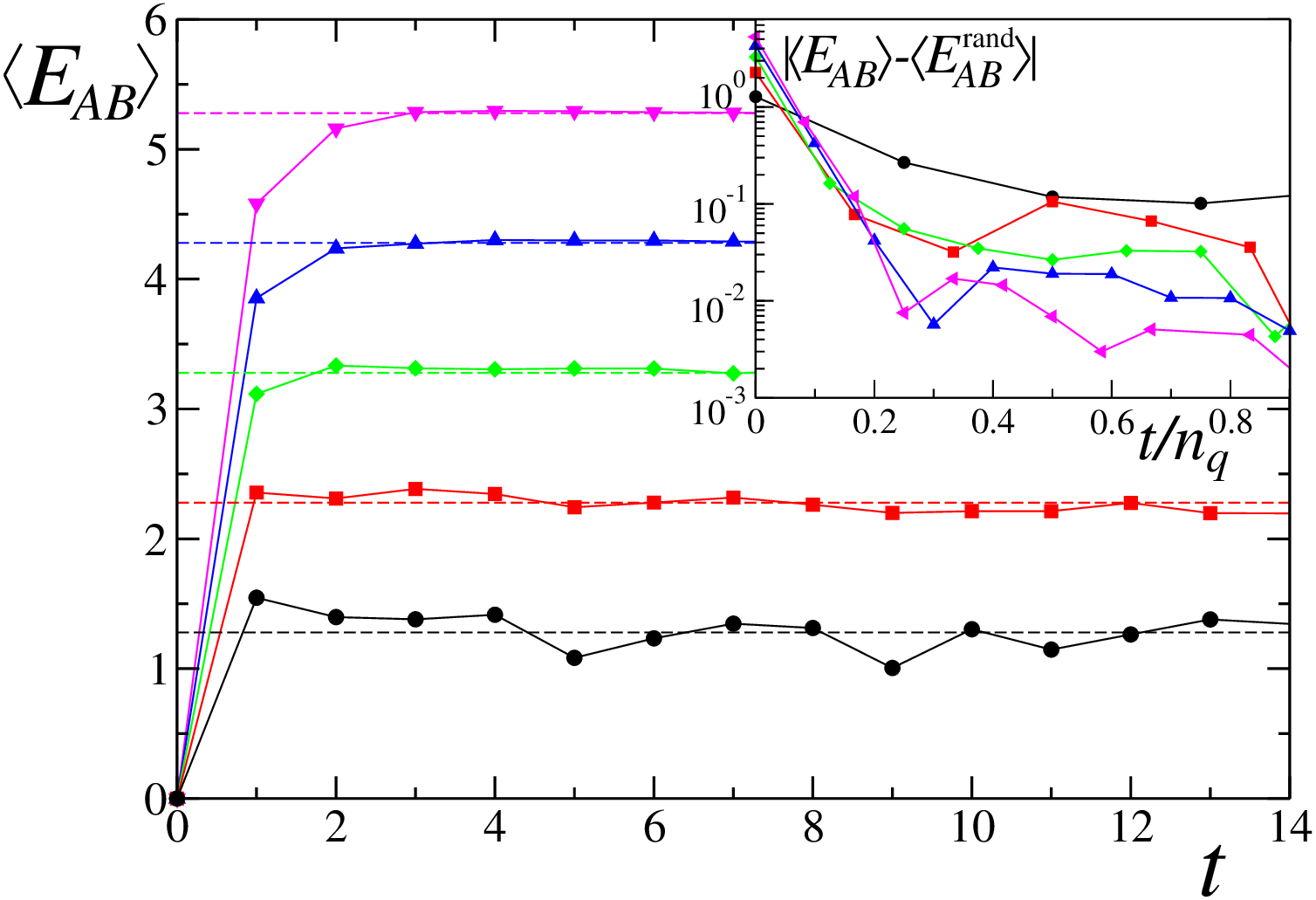}
    \caption{(color online) Time evolution of the average 
      bipartite entanglement of a quantum state,
      starting from a state of the computational basis
      (eigenstate of the momentum operator $\hat{n}$),
      and recursively applying the quantum sawtooth map~\eqref{eq:quantmap} 
      at $K=1.5$ and, from bottom to top, $n_q= 4, 6, 8, 10, 12$.
%      $n_q=$ 4 (circles), 6 (squares), 8 (diamonds),
%      10 (triangles up), and 12 (triangles down).
      Dashed lines show the theoretical values of
      Eq.~\eqref{eq:entpure}.
      Inset: convergence of $\langle E_{AB} \rangle (t)$
      to the asymptotic value
      $\langle E^{\, {\rm rand}}_{AB} \rangle$.}
%      in Eq.~\eqref{eq:entpure}; time axis is rescaled
%      with $1/n_q$.}
    \label{fig:EntGen}
  \end{center}
\end{figure}

\paragraph{Multipartite entanglement generation.}
We first compute $\langle E_{AB} \rangle$ as a function 
of the number $t$ of iterations of map~\eqref{eq:quantmap}.
Numerical data in Fig.~\ref{fig:EntGen} exhibit a fast convergence,
within a few kicks, of this quantity to the value 
\beq
\langle E^{\, {\rm rand}}_{AB} \rangle= 
\frac{n_q}{2} - \frac{1}{2 \ln 2} 
\label{eq:entpure}
\eeq
expected for a random state~\cite{page93}.
Precisely, %as shown in the inset of Fig.~\ref{fig:EntGen},
$\langle E_{AB} \rangle$ converges exponentially fast to
$\langle E^{\,{\rm rand}}_{AB} \rangle$, with the time scale 
for convergence $\propto n_q$ (see inset of Fig.~\ref{fig:EntGen}).
Therefore, the average entanglement 
content of a true random state is reached to a fixed accuracy 
within $O(n_q)$ map iterations, namely $O(n_q^3)$  
quantum gates.
We stress that in our case a deterministic map, 
instead of random one- and two-qubit 
gates as in Ref.~\cite{plenio}, 
is implemented. Of course, since the overall Hilbert space 
is finite, the above exponential decay in a deterministic map is possible
only up to a finite time and the maximal accuracy drops exponentially
with the number of qubits.
We also note that, due to the quantum chaos regime, 
properties of the generated pseudo-random state do not depend
on initial conditions, whose characteristics may even be very different
from it (e.g., in simulations of Fig.~\ref{fig:EntGen},
we start from completely disentangled states).

%%%%%%%%%%%%%%%%%%%%%%%%%%%%%
%%%%%%%%%%%%%%%%%%%%%%%%%%%%%

\begin{figure}[!ht]
  \begin{center}
    \includegraphics[scale=0.34]{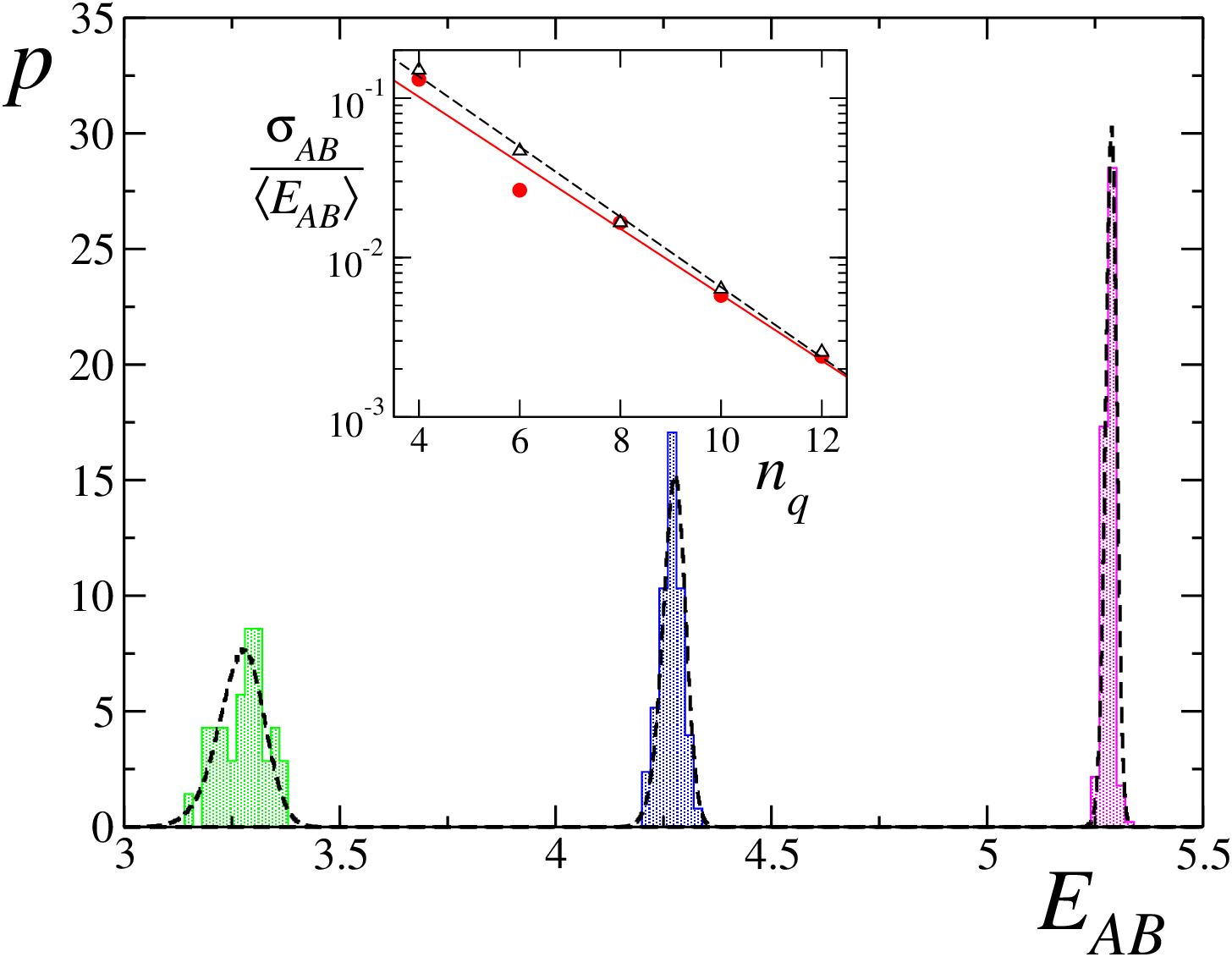}
    \caption{(color online) Probability density 
      function of the bipartite von Neumann
      entropy over all balanced bipartitions for the state $\ket{\psi_t}$,
      after $30$ iterations of map~\eqref{eq:quantmap} at $K=1.5$.
      Various histograms are for different numbers of qubits:
      from left to right $n_q = 8, 10, 12$;
      dashed curves show the corresponding probabilities for random states.
      Inset: relative standard deviation
      $\sigma_{AB} / \langle E_{AB} \rangle$ as a function of $n_q$
      (full circles) and best exponential fit
      $\sigma_{AB} / \langle E_{AB} \rangle \sim e^{- 0.48 \, n_q}$
      (continuous line); data and best exponential fit
      $\sigma_{AB} / \langle E_{AB} \rangle \sim e^{- n_q / 2}$
      for random states are also shown (empty triangles, dashed line).}
    \label{fig:Isto_eps0}
  \end{center}
\end{figure}

As discussed above, multipartite entanglement should 
generally be described in terms of a function, rather than by 
a single number. We therefore show in 
Fig~\ref{fig:Isto_eps0} the probability density function 
$p(E_{AB})$ for the entanglement of all possible balanced
bipartitions of the state $\ket{\psi_{t=30}}$.
This function is sharply peaked around 
$\langle E^{\,{\rm rand}}_{AB}\rangle$,
with a relative standard deviation $\sigma_{AB} / \left< E_{AB} \right>$
that drops exponentially with $n_q$ (see 
the inset of Fig.~\ref{fig:Isto_eps0})
and is very small ($\sim 0.1$) already at $n_q=4$.
For this reason, we can conclude that multipartite entanglement is 
large and that it is reasonable to use the first moment 
$\langle E_{AB} \rangle$ of $p(E_{AB})$ for its characterization.
We have also calculated the corresponding probability
densities for random states (dashed curves in Fig.~\ref{fig:Isto_eps0});
their average values and variances are in agreement with the values
obtained from states generated by the sawtooth map.
The fact that for random states the distribution $p(E_{AB})$ is 
peaked around a mean value close to the maximum achievable value 
$E_{AB}^{\rm max}=n_q/2$ is a manifestation of the ``concentration
of measure'' phenomenon in a multi-dimensional Hilbert space
\cite{zyczkowski,winter}.

%%%%%%%%%%%%%%%%%%%%%%%%%%%%%
%%%%%%%%%%%%%%%%%%%%%%%%%%%%%

\paragraph{Stability of multipartite entanglement.} 
In order to assess the physical significance of the 
generated multipartite entanglement, it is crucial to 
study its stability when realistic noise is taken into account.
Hereafter we model quantum noise by means of unitary noisy gates,
that result from an imperfect control of the quantum computer
hardware~\cite{cirac95}. We follow the noise model of Ref.~\cite{rossini04}.
One-qubit gates can be seen as rotations of the Bloch sphere
about some fixed axis; we assume that unitary errors slightly tilt
the direction of this axis by a random amount.
Two-qubit controlled-phase shift gates are diagonal in the
computational basis; we consider unitary perturbations by
adding random small extra phases on all the 
computational basis states.
Hereafter we assume that each noise parameter $\eps_i$ is randomly
and uniformly distributed in the interval $[-\eps, +\eps]$;
errors affecting different quantum gates are also supposed to be
completely uncorrelated: every time we apply a noisy gate, noise
parameters randomly fluctuate in the (fixed) interval $[-\eps, +\eps]$.

Starting from a given initial state $\ket{\psi_0}$,
the quantum algorithm for simulating the sawtooth
map in presence of unitary noise gives an output state
$\ket{\psi_{\eps_I,t}}$ that differs from
the ideal output $\ket{\psi_t}$. Here 
$\eps_I=(\eps_1,\eps_2,...,\eps_{n_d})$ stands for 
all the $n_d$ noise parameters 
$\eps_i$, that vary upon the specific noise configuration
($n_d$ is proportional to the number of gates).
Since we do not have any a priori knowledge of the 
particular values taken by the parameters $\eps_i$,
the expectation value of any observable $A$ for our
$n_q$-qubit system will be given by 
${\rm Tr} [\rho_{\eps,t} A]$, where the density matrix
$\rho_{\eps,t}$ is obtained after averaging over noise:
\beq
\rho_{\eps,t} = 
\left(\frac{1}{2\eps}\right)^{n_d} 
\int d \eps_I
\ket{\psi_{\eps_I,t}} \bra{\psi_{\eps_I,t}} \, .
\label{eq:rhomatr}
\eeq
The integration over $\eps_I$ is estimated numerically
by summing over $\mathcal{N}$ random realizations of noise,
with a statistical error vanishing in the limit $\mathcal{N}\to \infty$.
The mixed state $\rho_{\eps}$ may also arise as a consequence
of non-unitary noise; in this case Eq.~\eqref{eq:rhomatr}
can also be seen as an unraveling of $\rho_{\eps}$ into stochastically
evolving pure states $\ket{\psi_{\eps_I}}$, each evolution
being known as a quantum trajectory~\cite{brun}.

We now focus on the entanglement content of $\rho_{\eps,t}$.
Unfortunately, for a generic mixed state of $n_q$ qubits,
a quantitative characterization of entanglement
is not known, neither unambiguous~\cite{plenio07}.
Anyway, it is possible to give numerically accessible 
lower and upper bounds
for the bipartite {\it distillable entanglement}
$E_{AB}^{(D)} (\rho_{\eps})$:
\beq
\max \left\{ S(\rho_{\eps,A}) - S(\rho_\eps), 0 \right\} \leq
E_{AB}^{(D)} (\rho_{\eps}) \leq \log_2 \| \rho_{\eps}^{T_B} \| \, ,
\label{eq:entbounds}
\eeq
where $\rho_{\eps,A} = {\rm Tr}_B (\rho_\eps)$ and
$\| \rho_{\eps}^{T_B} \| \equiv {\rm Tr}
\sqrt{(\rho_\eps^{T_B})^\dagger \, \rho_{\eps}^{T_B}}$
denotes the trace norm of the partial transpose
of $\rho_{\eps}$ with respect to party $B$.

\begin{figure}%[!ht]
  \begin{center}
    \includegraphics[scale=0.31]{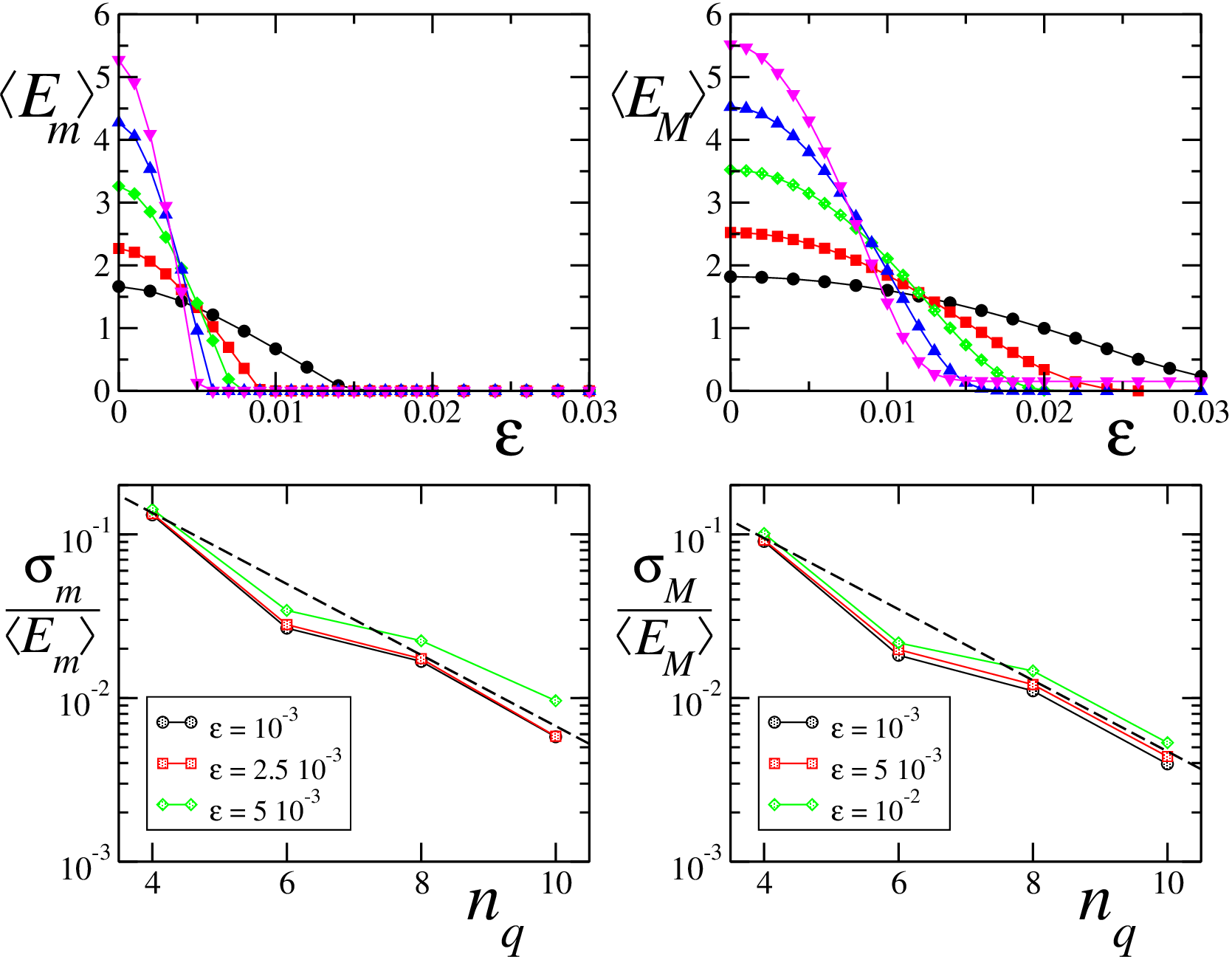}
    \caption{(color online) 
      Upper graphs: lower $\langle E_m \rangle$ (left panel) 
      and upper bound $\langle E_M \rangle$
      (right panel) for the distillable entanglement
      as a function of the noise strength at time $t=30$.
      Various curves stand for different numbers of qubits:
      $n_q=$ 4 (circles), 6 (squares), 8 (diamonds),
      10 (triangles up), and 12 (triangles down).
      Lower graphs: relative standard deviation of the probability density
      function for distillable entanglement over all balanced bipartitions
      as a function of $n_q$, for different noise strengths $\eps$.
      Dashed lines show a behavior
      $\sigma / \left< E \right> \sim e^{-n_q/2}$ and are
      plotted as guidelines.}
    \label{fig:Ent_dest_nqvar}
  \end{center}
\end{figure}

In practice, we simulate the quantum algorithm for the
quantum sawtooth map in the chaotic regime 
with noisy gates and evaluate
the two bounds in Eq.~\eqref{eq:entbounds} for the distillable
entanglement of the mixed state $\rho_{\eps,t}$,
obtained after averaging over $\mathcal{N}$ noise realizations.
A satisfactory convergence for the lower and the upper bound 
is obtained after $\mathcal{N}\sim \sqrt{N}$ and
$\mathcal{N}\sim N$ noise realizations, respectively. 
In Fig.~\ref{fig:Ent_dest_nqvar}, upper panels, we plot
the first moment of the lower ($E_m$) and the upper ($E_M$)
bound for the distillable entanglement 
as a function of the imperfection strength.
The various curves are for different numbers $n_q$ of qubits;
$\mathcal{N}$ depends on $n_q$ and is large enough 
to obtain negligible statistical errors (smaller than the size
of the symbols).
In the lower panels of Fig.~\ref{fig:Ent_dest_nqvar} we
show the relative standard deviation of the probability density
function (over all balanced bipartitions) for the distillable entanglement. 
Like for pure states, we notice an exponential drop with $n_q$;
the distribution width slightly broadens when increasing
imperfection strength $\eps$. 
We can therefore conclude that an average value of the bipartite
distillable entanglement close to the ideal case $\eps=0$ 
implies that multipartite entanglement is stable.

%%%%%%%%%%%%%%%%%%%%%%%%%%%%%
%%%%%%%%%%%%%%%%%%%%%%%%%%%%%

\begin{figure}%[!ht]
  \begin{center}
    \includegraphics[scale=0.31]{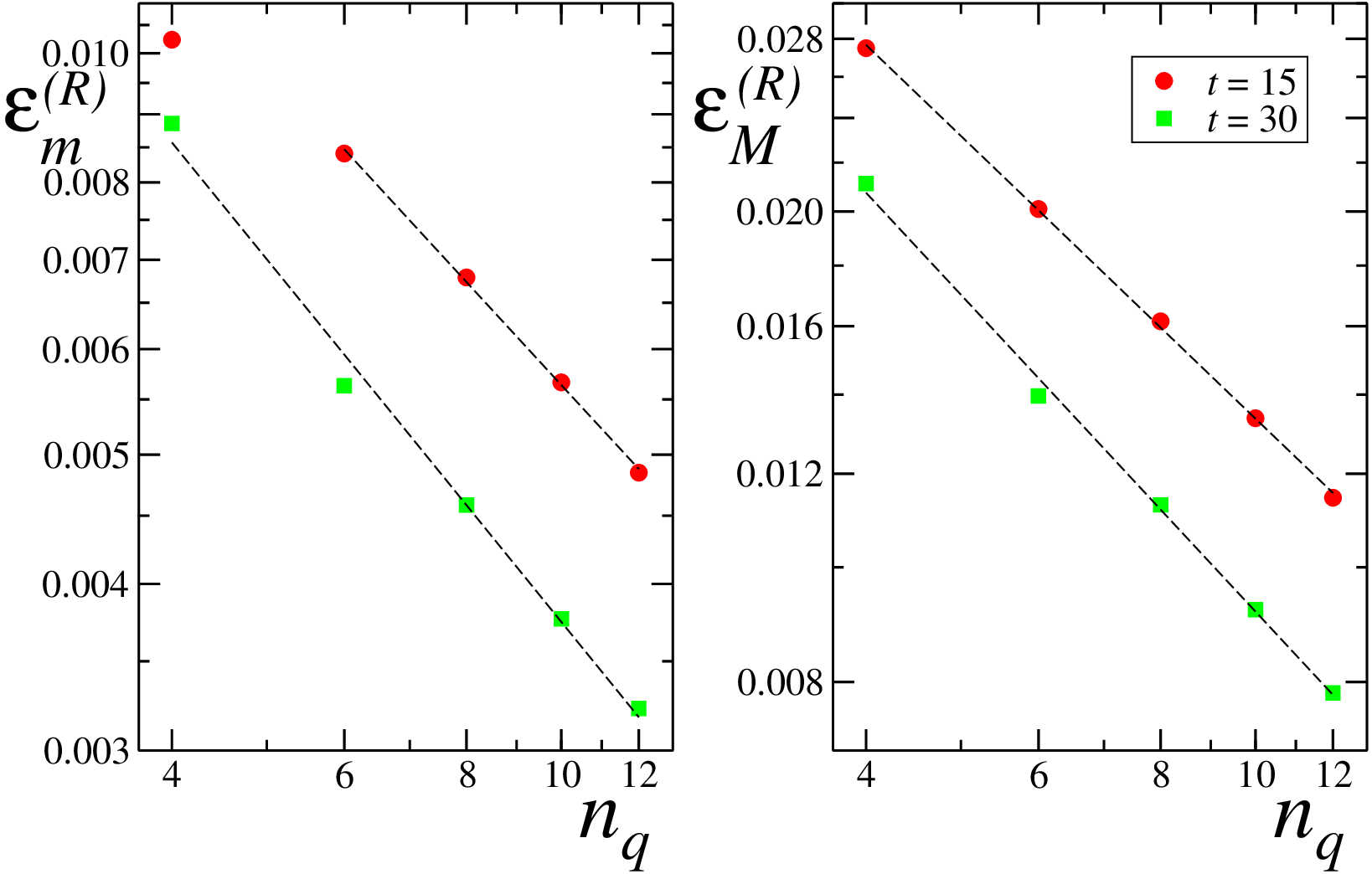}
    \caption{(color online) 
      Perturbation strength at which the bounds of multipartite
      entanglement halve (lower bound on the left panel, upper bound
      on the right panel), as a function of the number of qubits.
%      at $t=15$ (circles) and $t=30$ (squares).
      Dashed lines are best power-law fits of
      numerical data: $\eps^{(R)} \sim n_q^{-0.79 \pm 0.01}$ at $t=15$,
      $\eps^{(R)} \sim n_q^{-0.9 \pm 0.01}$ at $t=30$, for both
      lower and upper bounds.}
    \label{fig:EScal_eps_nq}
  \end{center}
\end{figure}

%\paragraph{Scaling of distillable entanglement decay with the system size.} 
In order to quantify the robustness of multipartite entanglement
with the system size, we define a perturbation strength threshold
$\eps^{(R)}$ at which the distillable entanglement bounds drop
by a given fraction, for instance to $1/2$, of their $\eps=0$ value, 
and analyze the behavior of $\eps^{(R)}$ as a function of
the number of qubits.
Numerical results are plotted in Fig.~\ref{fig:EScal_eps_nq};
both for lower and upper bounds we obtain a power-law scaling
close to
\beq
\eps^{(R)} \sim 1/n_q \, .
\label{eq:epsscaling}
\eeq

%
%%%%%%%%%%%%%%%%%%%%%%%%%%%%%
%%%%%%%%%%%%%%%%%%%%%%%%%%%%%
%

It is possible to give a semi-analytical proof of the
scaling~\eqref{eq:epsscaling} for the lower bound measure,
that is based on the quantum Fano inequality~\cite{schumacher96},
which relates the entropy $S(\rho_{\eps})$ to the fidelity 
$F = \langle \psi_t \vert \rho_{\eps,t}
\vert \psi_t \rangle$:
%\beq
$S(\rho_{\eps})\,\lesssim\, h(F) + (1 - F) \, \log_2 (N^2 - 1)\, ,$
%\label{eq:qfano}
%\eeq
where $h(x) = -x \log_2 (x) - (1-x) \log_2 (1-x)$ is the binary 
Shannon entropy.
Since $F \simeq e^{- \gamma \eps^2 n_g t}$~\cite{rossini04,bettelli04},
with $\gamma \sim 0.28$ and $n_g = 3 n_q^2 + n_q$ being the number of
gates required for each map step, we obtain, for $\eps^{2} n_g t \ll 1$,
\beq
S (\rho_{\eps}) \, \le \, \gamma \eps^2 n_g t \left[
 - \log_2 ( \gamma \eps^2 n_g t) + 2 n_q + \frac{1}{\ln 2} \right].
\label{eq:entrofano}
\eeq
For sufficiently large systems the second term dominates (for $n_q =12$
qubits, $t=30$ and $\eps \sim 5 \times 10^{-3}$ the other terms
are suppressed by a factor $\sim 1/ 10$) and, to a first
approximation, we can only retain it.
On the other hand, an estimate of the reduced entropy $S(\rho_A)$
is given by the bipartite entropy~\eqref{eq:entpure}
of a pure random state~\cite{page93}.
Therefore, from Eq.~\eqref{eq:entbounds} we obtain the following
expression for the lower bound of the distillable entanglement:
\beq
E_{AB}^{(D)} (\rho_{\eps}) \, \ge \,
\frac{n_q}{2} - \frac{1}{2 \ln 2} - 6 \gamma n_q^3 \eps^2 t \,.
\label{eq:fanoscaling}
\eeq
From the threshold definition
$E_{AB}^{(D)} (\rho_{\eps^{(R)}}) = \frac{1}{2} E_{AB}^{(D)} (\rho_{0})
=\frac{1}{2}S(\rho_A)$
we get the scaling~\eqref{eq:epsscaling}, that is valid
when $n_q \gg 1$:
$\eps^{(R)}_m \sim 1/\sqrt{24 \, \gamma \, n_q^2 \, t}$.
Notice that, for small systems as the ones that can be numerically
simulated (see data in Fig.~\ref{fig:EScal_eps_nq}),
the first term of Eq.~\eqref{eq:entrofano} may introduce remarkable
logarithmic deviations from the asymptotic power-law behavior.
At any rate, the scaling derived from Eq.~\eqref{eq:fanoscaling} is 
in good agreement with our numerical data, and also reproduces 
the prefactor in front of the power-law decay 
\eqref{eq:epsscaling} up to a factor of two.

\paragraph{Conclusions.}
We have shown that quantum chaotic maps provide 
a convenient tool to efficiently generate {\it in a robust way} the 
large amount of multipartite entanglement close to that expected 
for truly random states.
This result may become of practical relevance, 
since prototypes of quantum computers simulating these
systems and, in particular, our specific model~\cite{emerson}
have been already experimentally put on using a three-qubit
NMR-based quantum processor~\cite{cory,emerson}.
The fact that distillable entanglement of 
balanced bipartitions remains almost maximal, 
up to a noise strength which drops only {\it polynomially}
with the number of qubits, supports the possibility that 
multipartite entanglement of a large number of qubits 
might be used as a real physical resource in quantum information
protocols. 

\begin{acknowledgments}
We acknowledge 
support by MIUR-PRIN and EC-EUROSQIP.
This work has been performed within the
``Quantum Information'' research program of Centro di Ricerca
Matematica ``Ennio de Giorgi'' of Scuola Normale Superiore.
\end{acknowledgments}

\end{document}